# RF magnetron sputtering deposition of multilayers optical filters for ultra-broadband applications with a large number of thin layers

*Maxime* Duris[1,*], *Bryan* Horcholle[1], *Cédric* Frilay[1], *Christophe* Labbé[1], *Xavier* Portier[1], *Philippe* Marie[1], *Sylvain* Duprey[1], *Franck* Lemarie[1], *Julien* Cardin[1]

[1]Laboratoire CIMAP, Normandie Université, ENSICAEN, UNICAEN, CNRS UMR6252, CEA, 14000 Caen, France

**Abstract.** We present recent achievement on manufacturing optical filter and multilayers done with two complementary RF magnetron sputtering approaches: deposition duration control and in situ optical reflectance monitoring. Those approaches were greatly improved thanks to ellipsometry and spectrophotometry cross-studies of optical refractive indexes of $Nb_2O_5$, $TiO_2$ and $SiO_2$ materials grown using two sputtering systems. At the same time, we conducted deposition studies of these three materials which have increased the manufacturing reliability and allowed us to consider developing complex optical multilayers with more than 100 layers.

## 1 INTRODUCTION

Thin film coatings are technologically very important for applications in different scientific domains and technology sectors such as optics, energy and microelectronics and scientific instrumentation, medicine, display, communication. The stacks of different thin film materials are the basis of optical systems, which range from ultra-wideband mirrors and filters to narrowband filters, but also more complex systems such as Bragg mirror waveguides [1]. The performance and properties of these coatings increase with the complexity of thin film stacks. In order to explore this complexity, we investigate the potentiality of growing optical coatings composed of many layers for broadband filtering application.

To produce optical filters, a pair of materials of low optical absorption and low chromatic dispersion are chosen. High (H) and low (L) refractive index materials are selected to have the greatest possible index contrast and are stacked to form a doublet. By this way, strong transmittances or reflectances are obtained with a reduced number of H&L refractive index doublets. The multilayer structure is typically designed by either analytical or numerical methods. The first design methods use analytical formulations of the optical filter's response to design it. The second design method consists in H material insertion in a L one (or reciprocal) by selection of optimal position as for example by a needle method [2] followed by further thickness optimization.

One of the main issues of growing multilayer structures with high number of layers is the successive deviation of thickness $e$ and optical constant $n$ and $k$ called avalanche of deviation [9]. To realize complex optical filter composed of large number of layers, we have investigated two different sputtering approaches, with the aim to quantify and reduce this accumulation of errors.

## 2 METHODS AND EXPERIMENTS

### 2.1 GROWTH OF STRUCTURE BY TWO METHODS

In this work, we realize thin films and multilayer coating, with two different sputtering approaches. For both deposition methods, the growth was made on p-type 250 µm-thick [001] 2" silicon (Si) substrates by confocal reactive magnetron co-sputtering in a $O_2$/Ar rich plasma at close deposition pressure values of $2.10^{-3}$ mbar and $3.10^{-3}$ mbar.

Three different low optical absorption metal-oxide materials were investigated: in the UV to NIR spectral range the high refractive index materials $TiO_2$ [3, 4] and $Nb_2O_5$ [5, 6] and the well-known low refractive index material $SiO_2$ [7, 8].

First, a RF magnetron sputtering coating machine (AJA Intl.inc.) was used with prior evaluation of deposition rate. The 2'' $Nb_2O_5$ target was sputtered in $O_2$/Ar mixture flux to 2/8 sccm and 10 sccm pure Argon flux for $SiO_2$. During deposition, a sample control was used to adjust the deposition time for each layer with ex-situ spectroscopic ellipsometry measurement.

Second, a RF magnetron sputtering coating machine (Elettrorava S.p.A.), with in situ Optical Monitoring System (Intellemetrics) was used. The 3'' Ti, $SiO_2$ and $Nb_2O_5$ target powers were taken in the range for 200 W to 350 W. The sputtering plasma are composed of 40 sccm Argon flux and 5 to 12 sccm oxygen flux. The in situ optical monitoring system (OMS) is in reflection configuration directly on grown sample at a monitoring wavelength $\lambda_{OMS}$ ranging from 550 nm to 1650 nm. To be able to make comparison with AJA Intl.inc. coating machine and Elettrorava S.p.A., we injected the same power density on the $Nb_2O_5$ targets or 6.66 watt/cm², we apply a similarly ratio of deposition gazes Argon (80%)/Oxygen (20%).

### 2.2 OPTICAL CHARACTERISATION

---

[*] Corresponding author: maxime.duris@ensicaen.fr

The accuracy of determination of the complex refractive index on the whole spectral range of the targeted application is mandatory prior to the design step. First, the real refractive index n(λ) and extinction factor k(λ) of the $Nb_2O_5$, $TiO_2$ and $SiO_2$ materials were determined by variable-angle spectroscopic ellipsometry (Horiba UVISEL) with angle 65°, 70° and 75° with on monolayers deposited on Si wafer. The n(λ) and k(λ) were determined by ellipsometry modelling. Second, the complex refractive index was also determined by measuring and modelling the specular reflectance spectra obtained by means of a Perkin Elmer Lambda 1050 UV-Vis/NIR spectrometers with an integrating sphere from 250 nm up to 2500 nm spectral range. For each layer, the Optical Monitoring System (OMS) read the absolute reflectance in function of deposition time to control the deposition duration at selected wavelength λ. This OMS is a system that records reflectance R over deposition time, monitors the turning point [14-15] and trigg a cut-off condition on the deposition process when necessary.

To describe the refractive index and extinction coefficient dispersion, the New Amorphous (NA) dispersion model [10], was used to describe n(λ) and k(λ) for $SiO_2$ and $TiO_2$ material for both coating approaches while the Tauc-Lorentz (TL) model was used for $Nb_2O_5$. Those dispersion models use several parameters of interest which are the energy band gap $E_g$, the second is the strength (amplitude) and energy (position) of absorption/extinction coefficient peaks and the broadening terms $\Gamma_i$. Experimental uncertainties were obtained by a Monte Carlo re-sampling technique [11] based on uncertainties on NA and TL models parameters obtained by prior fitting with Horiba software.

## 4 RESULTS AND DISCUSSION

### 4.1 TiO$_2$, Nb$_2$O$_5$ AND SiO$_2$ MONOLAYERS

The first step consists in producing $TiO_2$, $Nb_2O_5$ and $SiO_2$ monolayers with both coating approach to study n(λ) and k(λ) dispersion (Figure 1 to Figure 3). The refractive index of materials is determined from experimental variable angle spectroscopic ellipsometry measurements.

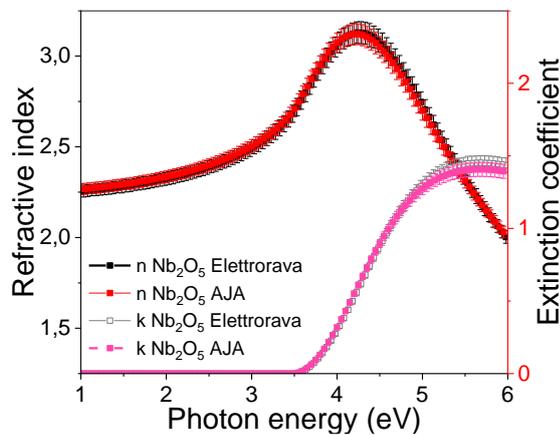

**Fig. 1.** *Refractive index and extinction coefficient dispersion of Nb$_2$O$_5$ Elettrorava and Nb$_2$O$_5$ AJA*

At first, we compared the $Nb_2O_5$ materials made in two RF magnetron sputtering deposition approaches with different geometry. The Figure 1 present a very good corresponds of optical characteristics for $Nb_2O_5$ deposited by this both approaches.

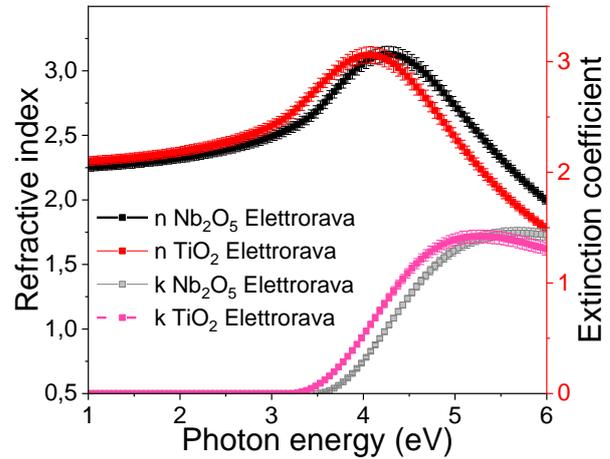

**Fig. 2.** *Refractive index and extinction coefficient dispersion of Nb$_2$O$_5$ Elettrorava and TiO$_2$ Elettrorava*

In the VIS-NIR part present to Figure 2, the refractive index of $TiO_2$ is slightly higher than the one of $Nb_2O_5$. Between 0.4 eV and 3.26 eV, the $n_{TiO2}$, varies from 2.285 ± 0.032 to 2.712 ± 0.046, while for $n_{Nb2O5}$ varies from 2.239 ± 0.0315 to 2.583 ± 0.0429. The extinction coefficient of $k_{TiO2}$ and $k_{Nb2O5}$ were found with a comparable evolution equal to 0 from 0.4 eV to 3.26 eV. $k_{TiO2}$ and $k_{Nb2O5}$ increase with slightly larger value for $TiO_2$ than $Nb_2O_5$ in the 3.26 eV- 4.75 eV energy range, while $k_{Nb2O5}$ becomes larger than $k_{TiO2}$ for energy higher than 4.75 eV.

Therefore, $TiO_2$ offer a higher contrast with $SiO_2$ than $Nb_2O_5$ in the UV-VIS-NIR for the realization of broadband optical filters but $TiO_2$ absorbs around 2.98 eV (420 nm) while $Nb_2O_5$ absorbs at 3.19 eV (380 nm).

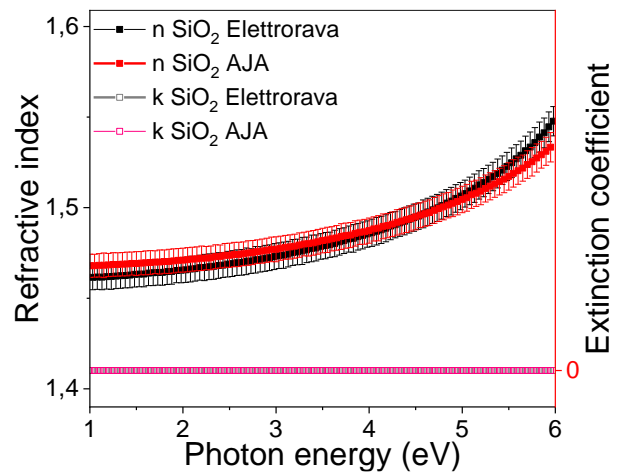

**Fig. 3.** *Refractive index and extinction coefficient dispersion of SiO$_2$ Elettrorava and SiO$_2$ AJA*

In Figure 3, considering the error bars, the refractive index n(λ) of $SiO_2$ layers produced by both deposition approaches (AJA and Elettrorava) varies similarly and monotonically from 1.4692 ± 0.005 at 0.4eV to 1.5435. ±

0.005 at 6 eV. For both deposition approaches, the extinction coefficient k(λ) was found equal to 0 on the whole spectral range. Despite the different deposition approaches, the uncertainties on n(λ) and k(λ) obtained by analysis of all Tauc-Lorentz parameter's uncertainties are also comparable. However, the $SiO_2$ monolayer realized with Elettrorava coating machine was found with an effective medium (EM) layer thick of about 2.4 nm while for $SiO_2$ obtained in AJA coating machine there were no EM layer found.

The observation of Figures 1 to 3 allows us to conclude that all materials are close to the standard values found in the literature. The behaviour of the band gap energy Eg of the 3 materials is presented on Figure 4. First, a small difference in the determination of $E_g$ for TL and NA dispersion models is observed. Second, the $E_g$ values of $SiO_2$ in AJA coating machine is more precisely determined than in Elettrorava coating machine probably due to the observed increase of roughness with the latter. Third, the $E_g$ values for $Nb_2O_5$ (Electrorava and AJA) are identical in accordance with the similarity of refractive index and extinction coefficient found in Figure 1.

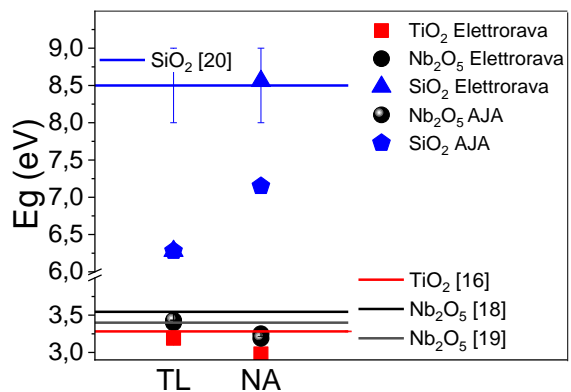

**Fig. 4.** *Energy band Gap $E_g$ as a function of Tauc-Lorentz (TL) and New amorphous (NA) model for $TiO_2$, $Nb_2O_5$ and $SiO_2$*

Atomic force microscopy (AFM) studies were done on three kind of materials. From AFM we determine the rms roughness of materials and we found no correlation length at the scale of interest (wavelength). Using the theory of light scattering by rough surface [12-13] the ratio of scattered light over total reflected was calculated taking into account our experimental roughness. For Electrorava $TiO_2$ monolayer the rms roughness was found equal to 1.4 nm, leading to the resulting light scattering ratio of 2.1 ± 0.4 %. For $Nb_2O_5$ monolayer realised in AJA, the EMA layer thickness was found equal to 3.4 nm, leading to the resulting light scattering ratio of 4.35 ± 0.45%. And for $SiO_2$ the rms roughness was found equal to 0.74 ± 0.23 nm leading to light scattering ratio equal to 1.75 ± 0.75 %. The contribution of roughness to optical losses due to surface scattering is thus quantified with values ranging from 1.75 to 4.35% depending on the deposited materials and on the deposition approach.

## 4.2 TiO₂/SiO₂ AND Nb₂O₅/SiO₂ MULTILAYERS

Study of two Broad Band Bragg Reflector (BBBR) composed of 36-layers of either $TiO_2$ and $SiO_2$ or $Nb_2O_5$ and $SiO_2$ were made. The both filters are an identical thickness structure. The spectral reflectance of these 36-layers BBBR is plotted in Figure 5 and 7 and presents a strong reflection from 450 nm to 1200 nm. We observe a good agreement between thickness adjustment spectra with Macleod for 36 layers $Nb_2O_5/SiO_2$ BBBR meanwhile a decreases of reflectance performance to 2.7% ± 0.7.

In fact, the χ² from thickness adjustment by Macleod with experimental spectra are 12.45 ± 2.25 for $Nb_2O_5$ / $SiO_2$ 36 layers BBBR.

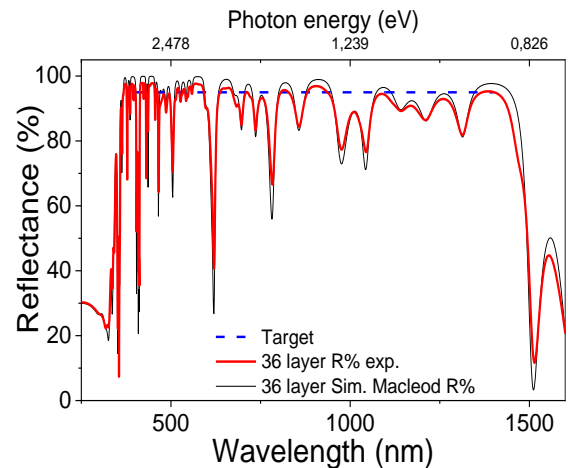

**Fig. 5** Reflectance of 36-layers Broadband Bragg Reflector $Nb_2O_5/SiO_2$ versus Macleod.

To compare optical performance of both 36-layer BBBR filter at central wavelength to 900 nm. The bandwidth Δλ/λ of 36-layers Broadband Bragg Reflector $Nb_2O_5/SiO_2$ are 1.18 ± 0.02 (or Δλ = 1066 ± 17 nm @ 900 nm).

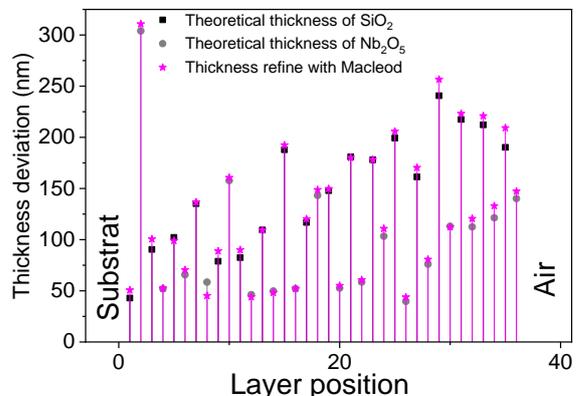

**Fig. 6** Macleod Absolute deviation between theoretical and simulated thicknesses of 36-layers broadband diffraction Bragg reflector $Nb_2O_5/SiO_2$

Figure 6, the absolute difference in thickness for 36 layers $Nb_2O_5/SiO_2$ BBBR is between 0.4% and 20%. The most important thickness deviation is on the layers with thickness lower than 70 nm.

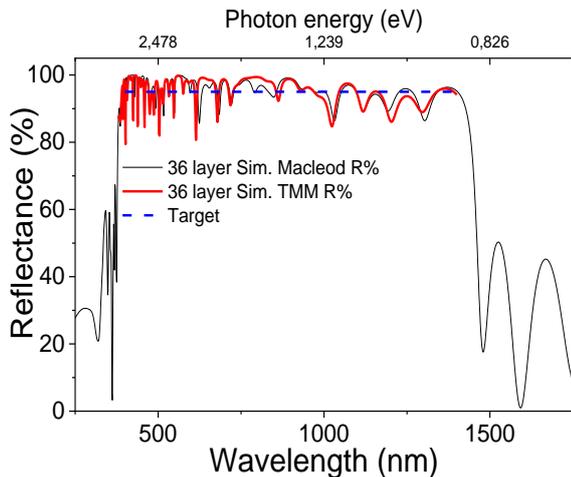

**Fig. 7** Reflectance of 36-layers Broadband Bragg Reflector TiO$_2$/SiO$_2$ Macleod and TMM simulation [17]

The next step in this study will be the realization of a 36-layer TiO$_2$ / SiO$_2$ BBBR optical filter with a same layer's structure as Nb$_2$O$_5$/SiO$_2$ and reflectance presented in figure 7. Finally, we will observe the comparison with TMM method [17] and Macleod simulation and will compare their optical performance in reflectance on the 400 nm to 1400 nm spectral range.

In conclusion, we managed to produce multilayer stacks with more than 36 layers. Thanks to the study of material's refractive index by means of New Amorphous and Tauc-Lorentz dispersion model we improved the characterization of material's optical properties $n(\lambda$ or eV) and $k(\lambda$ or eV). The high quality of three dielectric materials whose optical parameters were found with good agreements with those reported in literature permit to developed and realised a 36-layer BBBR with Nb$_2$O$_5$/SiO$_2$ and TiO$_2$/SiO$_2$ couples. This paves the way to ultra-broadband coatings with up to 100 layers, which may be use for different kinds of applications.